\RequirePackage{fix-cm}
\documentclass{svjour3}                     
\smartqed  
\usepackage{graphicx}
\usepackage{xcolor}
\graphicspath{ {./Figures_1/} }
\usepackage[utf8]{inputenc}
\usepackage[T1]{fontenc}
\usepackage{lmodern}
\usepackage{amsmath,amsfonts,amssymb}
\usepackage{geometry}
\usepackage[parfill]{parskip}
\usepackage{amssymb}
\usepackage{epstopdf}
\usepackage{color}
\usepackage{fancyhdr}
\usepackage{enumerate}
\usepackage{lastpage}
\usepackage{float}
\usepackage{gensymb}
\usepackage{siunitx}
\usepackage{comment}
\usepackage{authblk} 
\usepackage{caption}
\usepackage[font=small,labelfont=bf]{caption}
\usepackage{natbib}
\usepackage{ulem}
\usepackage{color,soul} 

\renewcommand{\figurename}{Fig.}
\journalname{Experiments in Fluids}
\begin{document}

\title{Characterization of a turbulent flow with independent variation of Mach and Reynolds numbers
}


\author{N. Manzano-Miura         \and
        D. Gloutak               \and
        G. P. Bewley
}


\institute{N. Manzano-Miura \at
              Cornell University, Sibley School of Mechanical and Aerospace Engineering, Ithaca, NY, USA \\
              \email{pnm24@cornell.edu}           
           \and
           D. Gloutak \at
              University of Colorado Boulder, Ann and HJ Smead Aerospace Engineering Sciences, Boulder, CO, USA
           \and
           G.P. Bewley \at
              Cornell University, Sibley School of Mechanical and Aerospace Engineering, Ithaca, NY, USA \\
}

\date{Received: date / Accepted: date}

\maketitle

\begin{abstract}
The Variable Density and Speed of Sound Vessel (VDSSV) 
produces subsonic turbulent flows 
that are compressible and in which turbulent fluctuations can be resolved at all scales with existing instrumentation including hot wires and particle tracking.
We realize this objective by looking at the flow of a heavy gas 
(sulfur hexafluoride SF\textsubscript{6}), 
with a speed of sound almost three times lower than for air. 
By switching between air and SF\textsubscript{6}, we isolate the influence of the turbulent Mach number (up to $M_t$ = 0.17) on turbulence statistics from the influences of changes in the Reynolds number (up to $R_{\lambda}$ = 1600), and boundary conditions, 
which we hold constant. 
A free shear flow is produced by a ducted fan, 
and we show that it behaves like a turbulent jet 
in that 
the mean velocity profiles approach self-similarity 
with increasing distance from the orifice (up to $x/D_f$ = 9). 
The jet responds like a 
compressible shear layer in that it spreads more slowly at higher Mach numbers (up to $M_j$ = 0.7) than at low Mach numbers. 
In contrast, the integral length scales 
and Kolmogorov constant of the turbulence are approximately 
invariant with respect to changes in either the Reynolds or Mach numbers. 
We briefly report on instrumentation under development 
that will extend the accessible Taylor-scale Reynolds and turbulent Mach numbers 
to 4000 and 0.3, respectively. 

\end{abstract}

\section{Introduction}
\label{sec:intro}

Hydrodynamic fluctuations in compressible turbulence play a determining role 
in engineered and natural flows \citep{Lele1994,Gatski2013}. 
A relatively long history of controlled experimentation
combined with theoretical analysis 
generates a detailed picture of incompressible turbulence  \citep[e.g.][]{Pope2001,davidson2012ten,sinhuber2015}, where 
the main parameter in the problem is the Reynolds number.
Though this picture is far from complete, 
it is much clearer than the one we have of compressible turbulence, where 
we have in addition to this parameter 
the Mach number 
and others whose roles have yet to be established definitively  \citep{danish2016influence,Donzis2019,Sabelnikov2021}.

In turbulence, fluctuations at small-scales depend sensitively on the Reynolds number in ways that can be difficult to distinguish from Mach number dependencies \citep{Jagannathan2016}. For instance, intermittency measured by the flatness of velocity derivatives increases as a power law of the Reynolds number, and is likely dependent on the Mach number in a way that depends on the flow geometry at the small-scale (be it vortex-like or shock-like). Due to the increased mathematical complexity 
and the larger parameter space that the turbulence explores,
there is less empirical or theoretical information available about compressible turbulence 
than about incompressible turbulence. 
For example, 
the energy dissipation rate, 
the small-scale structure of the flow, 
the shape of the energy spectrum, 
the translation between compressible forcing in simulations and realistic forcing in experiments, 
and the dependence of these features on the parameters in the problem are unknown. 

Although the work on free shear flows has clarified the main mechanisms by which compressibility dampens turbulent mixing, 
recent computer simulations suggest profound changes in the structure of compressible turbulence relative to incompressible turbulence, such as a sign change in the skewness of the distribution of pressure fluctuations or a rapid increase in the dilatational dissipation rate with increasing Mach number \citep{Jagannathan2016}. Moreover, recent literature discusses alternative mixing inhibition mechanisms that include the changing role of pressure and the influence of the orientation of disturbances at high Mach numbers \citep{karimi2016suppression,karimi2017influence}.
Even at subsonic mean speeds, 
turbulent fluctuations can be fast enough 
relative to the speed of sound 
that regions of local compression and expansion appear in the flow, 
albeit less frequently than at higher speeds \citep{Jagannathan2016,Wang2018}. 
The data needed to generate, validate 
and improve compressible turbulence models \citep{Georgiadis2014,quadros2016turbulent} 
are scarce due to difficulties in generating 
compressible flows with negligible density gradients (unlike in \citet{charonko2017variable} where variable density effects are characterized) in which turbulent fluctuations can be fully resolved.
In laboratory experiments the turbulent scales of motion are too small or too fast for most instrumentation. 
In contrast, in nature the conditions cannot be controlled. 

The wide range of applications in which 
compressible turbulence appears suggests a 
broad utility for an understanding of its universal aspects. 
In hypersonic boundary layers, for instance, 
not only the mean flow but also the fluctuations are compressible
\citep{Owen1975,Spina1994,Williams2018}. 
In scramjets and radial detonation engines 
combustion occurs at supersonic speeds, 
which translates to short residence times for the fuel and oxidizer in the combustor 
\citep{Urzay2018,Ladeinde2021}. 
In commercial and military aviation, 
the study of turbulence and its radiated sound 
have guided jet noise control strategies \citep[e.g.][]{bodony2008current,Jordan2013}. 
In astrophysical flows, 
star formation is slowed by supersonic turbulence \citep{Low2004,Federrath2013}. 
The early Universe has been modeled as a turbulent fluid that experiences turbulent mass fluctuations 
over physical scales that extend to tens of megaparsecs \citep{Shandarin1989}. 
These large-scale structures are associated with supersonic flows
that generate cosmological shocks in the intergalactic medium \citep{Ryu2008}. 

One of the few universal findings about compressible turbulent flows is that free shear layers grow more slowly at high Mach numbers than at low Mach numbers \citep{Gatski2013}. We briefly review these findings with a focus on turbulent jet experiments, since they resemble the experiments we performed.  
\citet{Schadow1990} reports Schlieren photography and total-pressure measurements and decreased spreading rates in coaxial jets at transonic Mach numbers ($0.25 < M_{c} < 2.25$, where $M_c = (U_{1}-U_{2})/(c_{1}-c_{2})$ is the convective Mach number and the subscripts $1$ and $2$ denote the faster and slower streams), 
and a strong dependence on the practical definitions of the jet width and spreading rate. 
The convective Mach number was modulated in part by varying the density ratio between the center and coaxial jets. The density ratio, whose effects are not studied here, also alter the development of compressible jets.
\citet{Samimy1993} 
finds that vortex generators at the jet orifice 
($0.3 < M_{j} < 1.81$, where $M_{j}=U_{j}/c$ is the jet Mach number
and $U_{j}$ is the jet speed at the orifice)
increase jet spreading rates 
in a way that depends on geometry but not on Mach number. 
\citet{Zaman1998,Zaman1999} 
also finds different behaviors for different nozzle geometries 
as well as decreased spreading and centerline velocity decay rates for round jets ($0.29 < M_{j} < 1.97$) above Mach one. 
\citet{Fleury2008} uses Particle Image Velocimetry (PIV) 
to find self-similar space-time correlations 
both off-axis and on-axis in subsonic isothermal round jets ($0.6 < M_{j} < 0.9$), 
as well as smaller integral scales at higher Mach numbers. 
\citet{Feng2016} studies subsonic but compressible annular shear layers in addition to supersonic ones,
($0.2 < M_{c} < 0.6$) 
and states that their slower growth at higher Mach numbers starts at lower $M_{c}$ than for planar mixing layers 
and is stronger at comparable $M_{c}$. 

Numerical simulations add detail to the experimental observation 
that compressible shear layers grow more slowly than incompressible ones. 
For instance, \citet{Sarkar1991}
implement a closure model and find reduced compressible shear layer growth rates 
consistent with experiments. 
\citet{Blaisdell1993} performs simulations of decaying isotropic, homogeneously sheared turbulence, 
and also finds a reduced growth rate of turbulence attributed to an increase in the dissipation rate accompanied by an energy transfer to internal energy by the pressure-dilatation correlation. 
\citet{FreundLeleMoin2000part1} reports direct numerical simulations (DNS) of annular mixing layers 
($0.1 < M_{c} < 1.8$, $0.2 < M_{j} < 3.5$, $M_{t}$ up to 0.8), 
which resemble the early development of a round jet, 
and explains suppressed growth rates by the budget of the streamwise component of Reynolds stresses -- 
whereas shear stresses and radial and azimuthal normal stresses are suppressed at high Mach numbers, the axial normal stresses stay the same, causing a shear stress anisotropy which decreases with Mach number. 
This latter study also identifies a decrease in transverse length scales due to an acoustic timescale limitation, 
which eventually causes large flow structures to deform faster than sound can propagate through them, disabling the formation of coherent eddies.
\citet{Sandberg2012} reports DNS of a compressible jet (0.46 < $M_{j}$ < 0.84) 
generated by a fully-developed pipe flow exiting into a low Mach number coflow, and finds that the self-similarity of coflowing jets breaks down for coflow values larger than about 40$\%$ of the jet speed. 
\citet{Arun2019} reports a DNS of high-speed mixing layers ($0.2<M_{c}<1.2$) 
and finds that with increasing $M_{c}$, an increase in vortex-dominated regions accompanies a reduction in shear layer growth rates. 

These findings are generally consistent with the observation that 
mixing layers spread more slowly at higher Mach numbers, 
an observation embodied in the Langley curve \citep{Slessor2000}. 
Taken together, the results also suggest an important dependence on 
the conditions of the flow at any given Mach number, 
and that different geometries produce different behaviors. 
The way turbulence changes with increasing Mach number 
needs to be explained and the mechanisms at 
work need to be separated from those associated with changes in the Reynolds number and geometry.
While the convective Mach number controls turbulent flow parameters such as the spreading rate of shear layers,
turbulence properties such as the flatness of velocity derivatives \citep{tang_antonia_djenidi_danaila_zhou_2018} depend on the Reynolds number, for instance, 
as do the slope of the energy spectrum \citep{praturi2019effect}, the intermittency exponent \citep{yakhot2018anomalous}, and the scaling exponents of structure functions \citep{iyer2020scaling}.

Incompressible turbulence within jets 
exhibits an approximate 
-5/3$^\mathrm{rds}$ power-law in energy spectra 
even in the near-field and at distances from the orifice smaller than those over which the jet is self-similar \citep{Fellouah2009}. 
This rapid development of the characteristic turbulence spectrum has also been observed in wakes and behind grids  \citep{Braza2006,valentevassilicos2012}. 
The spectrum may be steeper for compressible flows, 
when shocks are dominant for instance, 
and even for $M_t$ as low as 0.1 \citep{Bertoglio2001,Donzis2013,Federrath2013,Wang2017,Wang2018}. 
Such a steeper spectrum was reported in \citet{Biagioni1999} 
in a hypersonic wind tunnel flow albeit at an axial distance less than one nozzle diameter, 
where a -11/3$^\mathrm{rds}$ power law was in agreement with the compressible spectrum predicted by an Eddy Damped Quasi-Normal Markovian (EDQNM) model but inconsistent with a cascading picture characteristic of fully developed turbulence. 

This paper reports the first local quantitative measurements in compressible turbulence in the Variable Density and Speed of Sound Vessel (VDSSV), along with details of its design principles and capabilities. We describe the turbulent jet flow produced within our facility and its evolution with Mach number to benchmark the conditions in which the turbulence develops.
We generate a compressible flow and 
report quantities including two-point correlations, 
turbulent Mach number profiles, 
Kolmogorov constants, 
spectra, 
and their evolution with Mach number. 
We then compare our results with previous work.

\section{Apparatus and Methods}

To raise the Mach number at constant Reynolds number 
and for fixed boundary conditions, 
we compared experiments performed in air 
with ones performed in SF\textsubscript{6}.
The chief advantage of SF\textsubscript{6} 
is that its speed of sound is approximately 
2.5 times lower than the one for air. 
In contrast to air-breathing facilities running at similar Reynolds and Mach numbers, SF\textsubscript{6} experiments operate at lower speeds that are easier to resolve temporally since timescales are set by the speed of sound at any given Mach number.
Similarly, the small scales of turbulence are larger and slower at a given Mach number which enables improved spatial resolution. Additionally, SF\textsubscript{6} experiments
generate lower forces that ease mechanical design, 
consume less power and so facilitate electrical design, 
and can be run continuously 
in contrast to blow-down or burst-disk experiments 
so that statistics better converge. 
Finally, 
measurements from high (SF\textsubscript{6}) and low (air) 
Mach number experiments 
can be made with the same probes 
operating at the same frequencies. 

An alternate strategy of 
raising the Mach number of a gas flow 
at constant Reynolds number and fixed geometry 
is to increase the flow speed 
while simultaneously decreasing the pressure 
at constant temperature. 
This is so since 
the Reynolds number is proportional to changes in the gas density, 
while the speed of sound and dynamic viscosity are insensitive to it. 
This strategy was employed in the 1950's in 
the Variable Density High Speed Cascade Wind Tunnel at the Deutsche Forschungsansfalt, 
in which the gas pressure was varied between 0.1 and 1\,bar 
and the mean flow Mach number between 0.2 and 1.1
in order to characterize the performance of compressor blade cascades \citep{Schlichting1956}. 
This strategy, however, 
comes at the cost of increasing the frequency of the fluctuations 
that need to be detected (in proportion to the Mach number), 
so that faster instruments are needed at higher Mach numbers. 
To compare high and low Mach number measurements of turbulent fluctuations therefore, 
the frequency response of the instruments needs to be understood \citep{hutchins2015direct}. 
Even with the improvements in measurement technology since \citet{Schlichting1956}, 
we need to bring down the speed of the flow while holding constant its Mach number. 

\subsection{Pressure vessel}

\begin{figure}
    \centering
    \includegraphics[width=0.45\textwidth, trim=4cm 4cm 4cm 4cm, clip]{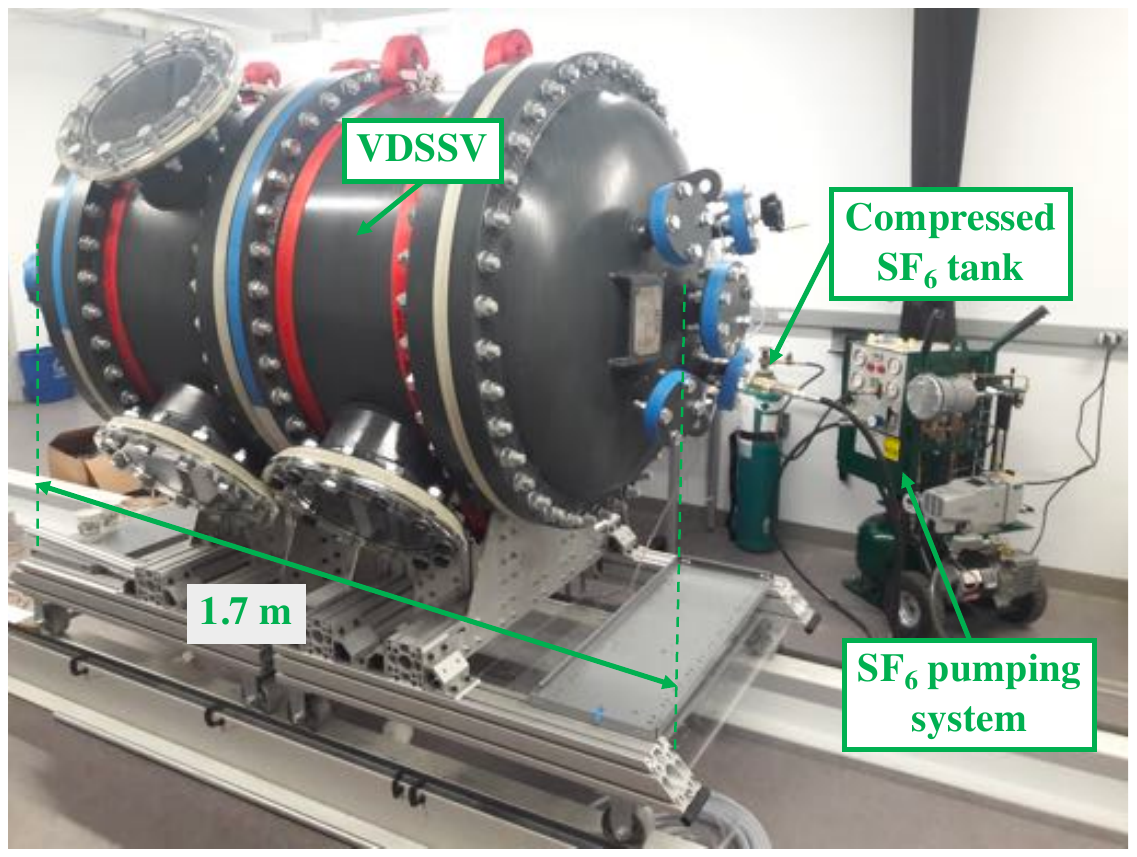}
    \includegraphics[width=0.45\textwidth, trim=3cm 4cm 3cm 4cm, clip]{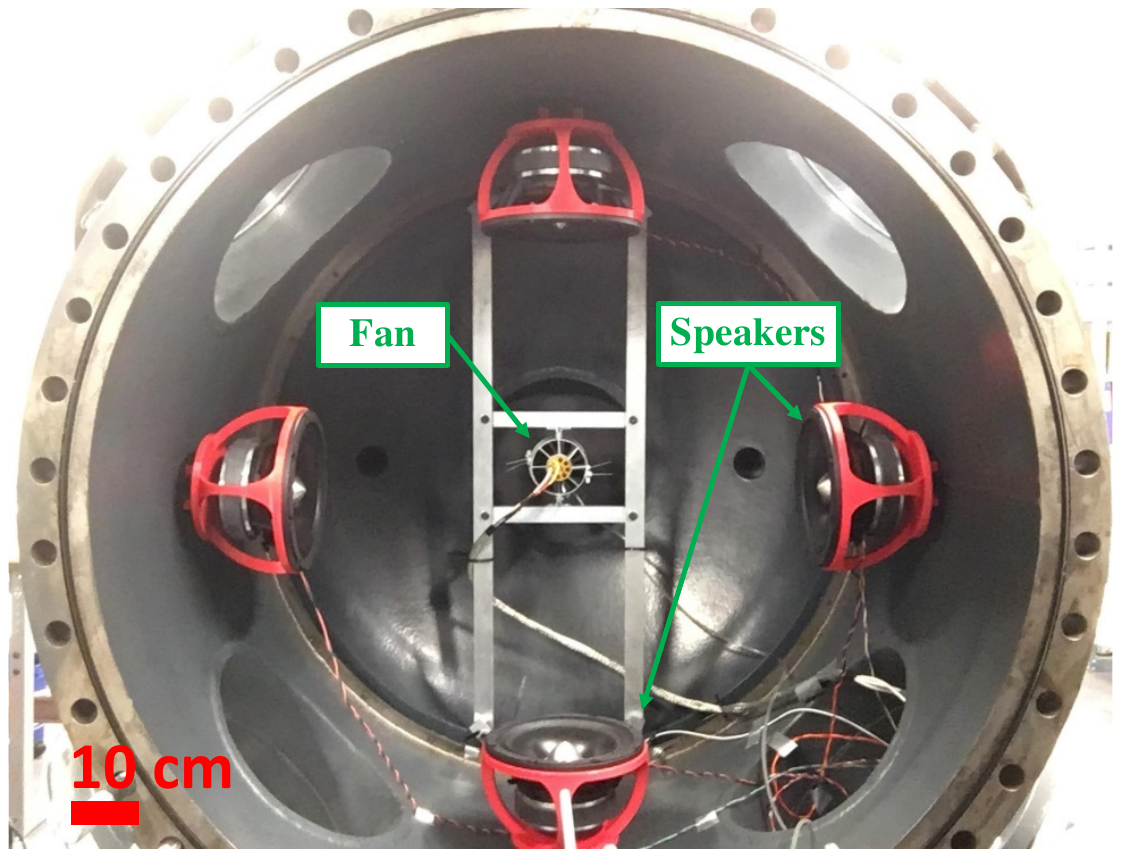}
    \caption{
    \textit{Top:} The facility includes the VDSSV, 
    a compressed SF\textsubscript{6} storage tank, 
    and the SF\textsubscript{6} gas-handling system. 
    Windows provide optical access through five 30\,cm portholes. 
    The VDSSV is mounted on carts and rails, 
    so that its two halves can be separated to access its interior. 
    \textit{Bottom:} A view upstream along the axis of the VDSSV, 
    once it has been separated along its midline, 
    shows the fan 
    that produces the turbulence characterized in this paper. 
    The red cages house loudspeakers for acoustic experiments not discussed here. 
    }
    \label{fig:1}
\end{figure}

We conducted experiments in a 16\,bar pressure vessel (Fig.~\ref{fig:1} 
and Fig.~\ref{fig:2}) 
consisting of two 0.6\,m long cylindrical sections with 
1\,m diameter and two rounded end caps made of hot-rolled steel. 
Each half of the vessel is mounted onto a cart that rolls on rails 
in order to pull the halves apart 
and to open the full cross section of the vessel to the experimenter. 
The length of the pressure vessel, now 1.7\,m, 
can be extended to 3\,m with the addition 
of annular sections between the two halves. Acrylic windows bolted to 30\,cm diameter 
portholes welded onto the shell 
provide optical access up to gas pressures of about 3\,bar \citep{Gloutak2018}. 
These acrylic windows can be exchanged with smaller sapphire windows (not shown) 
to extend this access up to 16\,bar. 

To run experiments with SF\textsubscript{6}, 
the VDSSV was first evacuated with a vacuum pump down to an absolute pressure of 0.01\,bar after ensuring that no leaks were present per \citet{Rottlander2016}, keeping the leak rate below 1.6$\times$10\textsuperscript{-1}\,mbar\,L/s. The facility was then charged to a set pressure from an external tank using a system of pumps and compressors (Enervac’s GRU-4 SF\textsubscript{6} Recovery Unit). 
After each experiment, the gas was stored and recycled. 
At a pressure of 0.165\,bar, the density of SF\textsubscript{6} 
is such that the kinematic viscosity, $\nu = \mu/\rho$, 
is the same as the one for air, 
and consequently the Reynolds numbers are matched. 

\begin{table}
\centering
\caption{Gas properties. 
$P$ is the static pressure, $\mu$ is the dynamic viscosity, $\gamma$ is the specific heat ratio, $c$ is the speed of sound, $T$ is the mean gas temperature, $\rho$ is the mean gas density. }
\label{tab:gas properties}       
\begin{tabular}{llllll}
\hline\noalign{\smallskip}
Gas & $P$ [bar] & $\mu|_{1 bar}$ [Pa $\cdot$ s] & $c$ [m/s] & $\rho$ [kg/m$^3$] & $T$ [K] \\
\noalign{\smallskip}\hline\noalign{\smallskip}
Air & 1.0 & 1.845 $\times 10^{-5}$  & 345  & 1.17 & 300\\
SF\textsubscript{6} & 0.165 & 1.549 $\times 10^{-5}$ & 136  & 0.97 & 300 \\
\noalign{\smallskip}\hline
\end{tabular}
\end{table}

\subsection{Turbulence generation}
\begin{figure}
    \centering
    \includegraphics[width=0.4\textwidth, trim=2.5cm 3.75cm 2.5cm 3.75cm, clip]{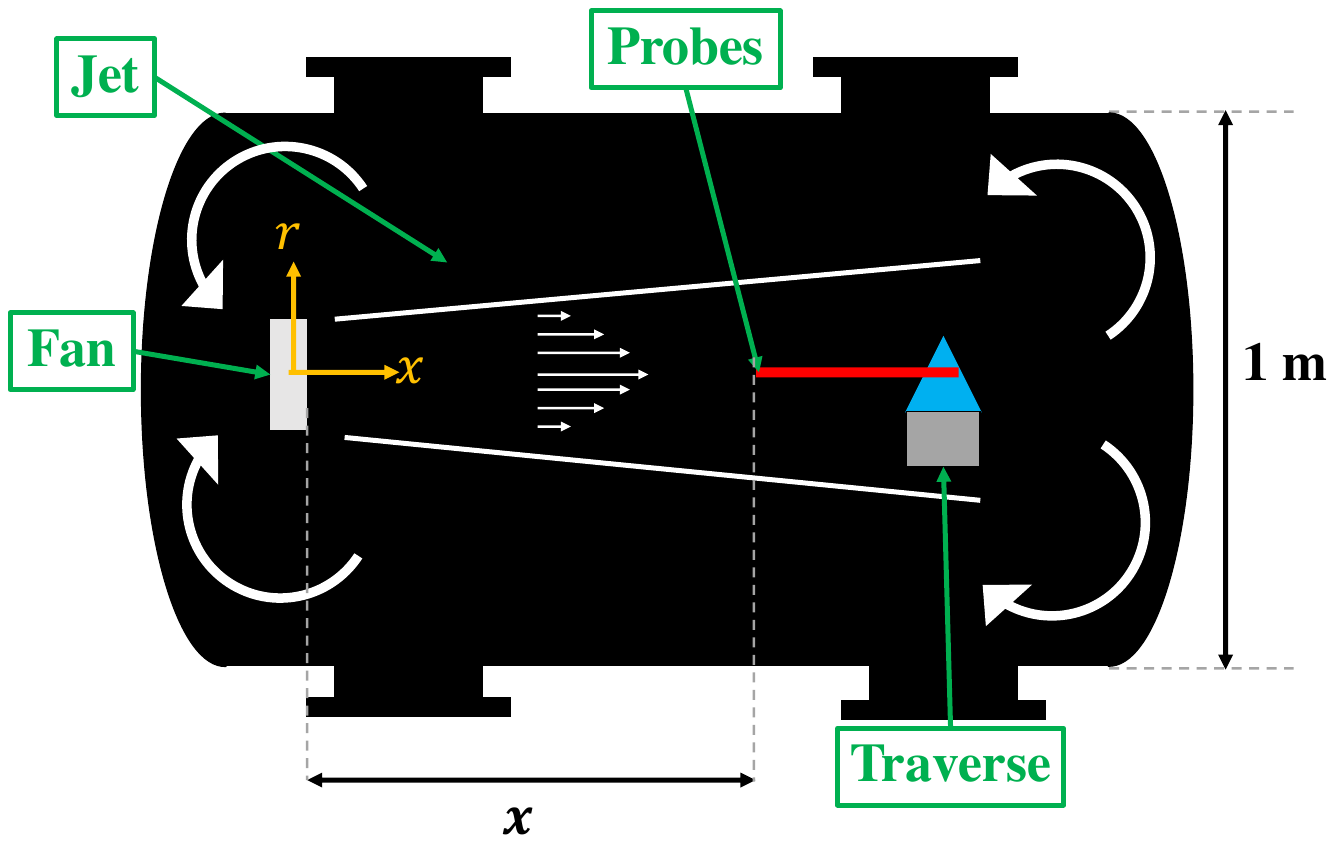}
    \caption{
    A sketch (not to scale) of the interior of the VDSSV 
    shows the configuration of the fan and probes. 
    The distance between the fan and the wall upstream of it is 25\,cm. 
    The distance along the axis of the fan from its outlet to the probe ($x$), 
    can be varied between 0.1 and 1.3\,m (1 and 15 fan diameters). 
    Orthogonal to the $x$-direction, $r$ is the radial distance outward from the centerline. 
    }
    \label{fig:2}
\end{figure}

A 12-blade ducted fan with a diameter $D_{f}=$ 90\,mm produced a turbulent jet which expanded 
down the centerline of the pressure vessel 
(see Fig.~\ref{fig:2}). In this sealed environment, the jet discharge fluid and the entrained external ambient fluid were identical.
The fan drew up to 2.4\,kW of electrical power,
part of which was converted into mechanical energy in the flow, 
and all of which was ultimately converted into heat. 
Fan exit flow velocities, measured by a Pitot tube, ranged from 10 to 90\,m/s, corresponding to jet Mach numbers up to approximately 0.7. This is well into the compressible regime, which is conventionally taken to prevail when $M_{j} > 0.3$.
When the turbulent jet impinges on the pressure vessel wall opposite to that of the fan, it generates a return flow near the outer walls of the pressure vessel (Fig.~\ref{fig:2}) , as in other turbulent jet experiments in enclosed facilities \citep[e.g.][]{Chanal2000}. The return flow is slower than the jet by a factor of about one hundred, which is the ratio of cross-sectional areas between the jet and the annulus around it, and is always subsonic.
From the mean rotation rate of a streamer in the jet, 
we estimated that the swirl number, $S = U_{\theta}/U_{j}$, 
which compares the the azimuthal to axial components of momentum, 
was smaller than 0.08 
over the full range of fan speeds. 
Swirl tends to increase jet spreading rates when it is larger than 0.6 
\citep{Gilchrist2005,vanierschot2008}, 
which is the opposite of the expected effect of compressibility. 

We classified our observations in terms of 
the Taylor microscale Reynolds number, $R_{\lambda} = {u^{\prime}\lambda}/{\nu}$ 
and the turbulent Mach number, $M_{t} = \sqrt{3} \, u^\prime / c$ \citep{Samtaney2001}. 
The factor of $3^{1/2}$ has its origin in the isotropic assumption that $u' \approx v' \approx w'$, 
which we retain though jet turbulence is anisotropic, 
particularly at its margins. 
As shown below, the jet comprises velocity fluctuations up to approximately 
$u^\prime = 15\,m/s$, corresponding to $R_{\lambda}$ up to 2000 and $M_{t}$ up to 0.2. 
\subsection{Diagnostic apparatus}
\label{sec:experimental method}

We used constant temperature anemometry (CTA) with hot-wire probes 
that resolved inertial-range statistics. 
The VDSSV produces a subsonic, compressible jet whose Kolmogorov scales can be as small as 10 microns. 
The probes were made of Wollaston wire, and measured 0.6 microns in diameter 
and about 120 microns in length, 
which corresponds to a spatial resolution of between 1 and 12 Kolmogorov scales. 
Hot wires measure a cooling rate 
due to heat transfer rate from the wire to the flow, 
which in this paper we interpret as a cooling velocity, $u$. 
In the compressible flow regime, 
hot wires respond to both mass flux and temperature fluctuations. 
The balance between these two types of fluctuations 
depends on the Mach number \citep{quadros2016kovasznay} 
and on the wire temperature. 
In the present experiment, the probes were operated at an overheat ratio of 1.4, 
so that the probes responded predominantly to mass flux fluctuations. 
Assuming that the Strong Reynolds Analogy holds \citep{Gatski2013}, 
density fluctuations were approximately 0.04\,$\%$ of the mean 
at an axial position of 9 fan diameters, where turbulence measurements are reported. 
Hence, density fluctuations were considered small enough compared to the velocity fluctuations, which at the jet center achieved values of 25\,$\%$ of the mean. 

Resolving diminutive scales is challenging not only due to the limited spatial resolution of modern probes, but also due to the proximity to the limit between the continuum regime and the slip-flow regime. Measurements in this regime were performed, for instance, by \citet{Kokmanian2019}, where a modified version of the state-of-the-art nanoscale thermal anemometry probes (NSTAPs) were used to measure turbulence at supersonic conditions, and found a linear relationship between Reynolds number and Nusselt number due to the closeness of the probe diameter to the mean free path of molecules (Knudsen number, $Kn ~ O(1)$). 
In our experiment, the mean free path in SF\textsubscript{6} is approximately 200\,nm, corresponding to a wire-diameter based Knudsen number of 0.33, and a micro-structure Knudsen number, based on the Kolmogorov scale, of 0.017. The former Knudsen indicates that our probes are not strictly operating on a continuum flow regime. 
The latter Knudsen number suggests that, at low speeds of sound, the smallest flow scales produced in the experiment are marginally within the continuum limit. In fact, the micro-structure Knudsen number increases linearly with $M_{t}$ and with the inverse square root of $R_{\lambda}$ \citep{Gatski2013}. Hence, in the limit of very low Reynolds numbers and very high Mach numbers, turbulent eddies may interfere with molecular motion \citep{tennekes2018first}. 
Nevertheless, we discard this possibility in our experiment, as it produces Reynolds numbers high enough to enable experimentation for a broad range of Mach numbers where the motion of gas molecules is in statistical equilibrium and molecular transport effects can be represented by transport coefficients such as the viscosity.

Hot-wire voltages were acquired at streamwise positions between ${x}/{D_{f}} =$ 1.9 and 9.1, 
which we call station 1 and station 2, and sampled with a digital acquisition card (16-bit NI-USB 6221) at a frequency, $f_{s}$, of 200\,kHz for approximately 10\,s (corresponding to about $1.7 \times 10^{4}$ to $1.7 \times 10^{5}$ integral time scales), and low-pass filtered at the Nyquist frequency, $f_{Nyq}$, of 100\,kHz to prevent aliasing. 

The probes were calibrated \textit{in situ} against a Pitot tube while varying the fan speed. 
We employed King's Law to relate mean voltages and speeds \citep{King1914}, 
along with a temperature correction \citep{bruun1996hot}. 
Along the axis of a jet, the velocity distribution is approximately 
Gaussian \citep{Anselmet1984}. 
\citet{Chanal2000} exploit this fact to determine a calibration 
from the differences between a measured hot-wire voltage distribution 
and a Gaussian distribution 
modified to account for probe behavior at near-zero speeds, such that 
\begin{equation}
    P(u) \propto u^{2} \exp \left( \frac{\left( u - U \right)^2}{2\sigma^{2}} \right) \label{eq:Assumed_PDF}
\end{equation}
where $u(x,t)$ is the flow speed, 
$U$ = $\langle u \rangle$ is its mean, 
and $\sigma^{2}$ = $\langle u^2 \rangle$ is its variance. 
We used this method to extend the calibration given by King's law 
to velocities beyond those for which we made Pitot tube measurements. 
The spatial structure of turbulence was investigated by invoking Taylor's Hypothesis of frozen turbulence \citep{taylor1937statistical}. 
We estimated the turbulent kinetic energy dissipation rate, $\epsilon$, 
from the inertial-range scaling of second-order velocity structure functions, particularly by taking the plateau value in $S_2 = C_2 (\epsilon r)^{2/3}$ and solving for $\epsilon$ \citep{Pearson2002}. 
Large scale turbulence statistics were computed as follows. 
The amplitude of velocity fluctuations was $u' = \langle u^{2} \rangle^{1/2}$.
The integral scale was taken as the area under the velocity autocorrelation function, 
$L = \int_{0}^{\infty} \! f(r) \, dr$ with $f(r)=\langle u(x+r)u(x) \rangle / \langle u^{2} \rangle$, 
where the tails for correlations below 0.1 
were calculated from integrals of exponential fits 
to the autocorrelation functions between values of $1/e$ and 0.1 
\citep{Bewley2012}. 
The Taylor microscale was computed as 
$\lambda = u^\prime \sqrt{{15\nu}/{\epsilon}}$, 
and the Kolmogorov length scale as $\eta = \left( {\nu^{3}}/{\epsilon} \right)^{1/4}$. 

\section{Results}
\label{sec: Results}
We survey mean flow characteristics as well as turbulence statistics 
in the turbulent jet 
and compare the data with those available in the literature. 
A description of the parameter space covered by the present experiment is followed by a characterization 
of the turbulent jet at stations 1 ($\approx 2D$) and 2 ($\approx 9D$), 
and their evolution with increasing Mach number, 
all of which anchor our analysis of second-order turbulence statistics such as autocorrelation functions and spectra. 

\subsection{Parameter Space}

\begin{figure}
\centering
	        \includegraphics[width=0.5\textwidth,trim=0cm 0cm 0cm 0cm, clip]{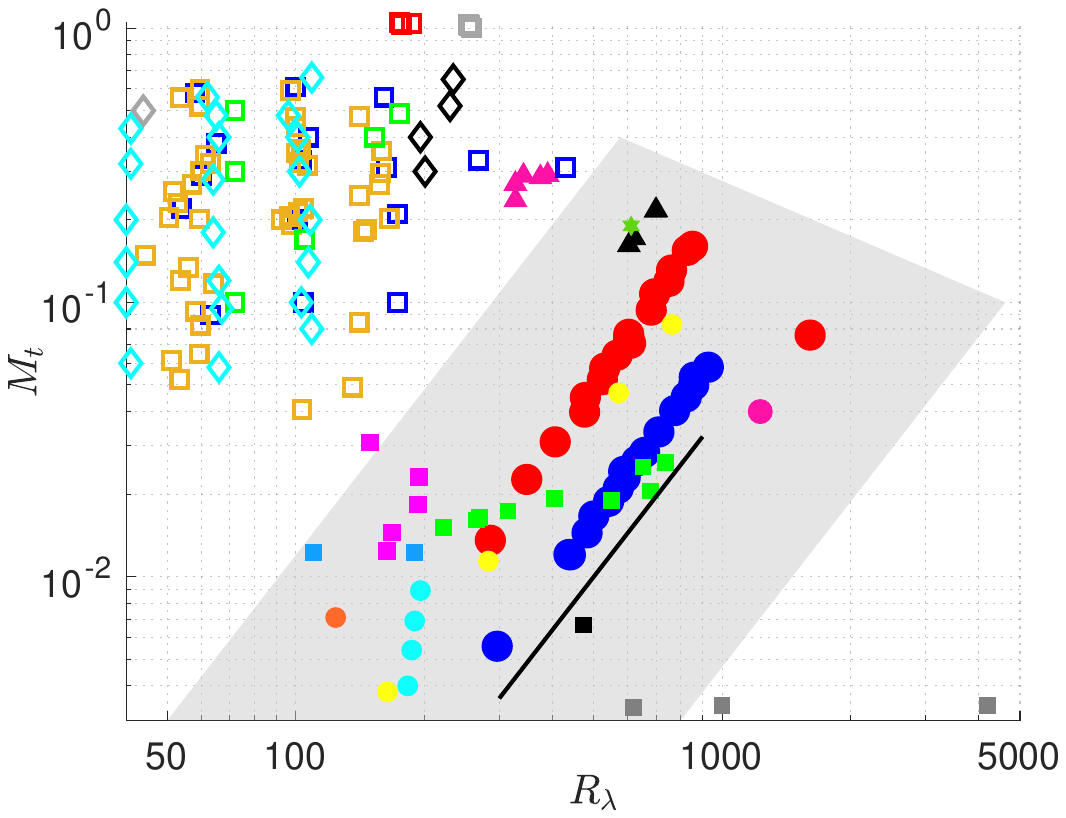}
	        \caption{
	        The $R_{\lambda}-M_{t}$ parameter space 
	        achieved in various experiments (closed symbols) 
	        and simulations (open symbols). 
	        Different shapes correspond to different flow type as follows: 
	        jets (circles), 
	        boundary layers (hexagrams), 
	        mixing layers (triangles), 
	        homogeneous shear (diamonds), and 
	        homogeneous isotropic flows (squares). Our data are shown as closed blue and red circles for air and SF\textsubscript{6}, respectively.
	        Other sources of experimental data are as follows.
	        Circles:
	        light blue \citep{Wygnanski1969}, 
	        orange \citep{Hussein1994}, 
	        magenta \citep{Biagioni1999}, 
	        yellow \citep{Narayanan2002}.
	        Hexagrams:
	        green \citep{Spina1987}. 
	        Triangles:
	        magenta \citep{Barre1994}, 
	        dark gray \citep{Bowersox1994}. 
	        Squares:
	        black \citep{Mydlarski1996}, 
	        light blue \citep{Zwart1997}, 
	        magenta \citep{Honkan1992}, 
	        light green \citep{BRIASSULIS2001}, 
	        gray \citep{Bodenschatz2014}.
	        An abundance of approximately incompressible turbulence experiments not shown 
	        lie below $M_t$ = 0.01 \citep[e.g.][]{sinhuber2015}. 
	        The DNS are as follows. 
	        Squares:
	        green \citep{Samtaney2001}, 
	        red \citep{Wang2011}, 
	        gray \citep{Wang2012}, 
	        dark blue \citep{Donzis2013},
	        orange \citep{Donzis2019}.  
	        Diamonds:
	        gray \citep{Blaisdell1993}, 
            light blue \citep{Chen2018}, 
	        black \citep{Wang2018}. 
		    }
		    \label{fig:3}
\end{figure}

The parameter space of the Reynolds number ($R_{\lambda}$) 
and turbulent Mach number ($M_{t}$) 
accessed in a representative collection of experiments and numerical simulations 
drawn from the literature 
is shown in Fig.~\ref{fig:3}. 
Data from simulations, most of which are DNS, span a range of turbulent Mach numbers up to approximately 1.0
and Taylor Reynolds numbers rarely exceeding 100. 
At higher Reynolds numbers, 
the simulation of compressible turbulence becomes prohibitive in terms of computational cost and the small-scale flow physics need to be modeled. Additionally, low Mach number simulations are difficult due to the separation of the acoustic and turbulence time scales. 

Previous experiments, in contrast to simulations, 
easily reach Reynolds numbers $R_{\lambda} > 100$, 
a regime in which turbulence is well developed in the sense that an inertial subrange emerges where turbulence is only weakly affected by boundaries and friction \citep[][]{Pope2001}. 
Attaining high Mach numbers in experiments poses challenges both from the power requirement and measurement resolution perspectives. 
Turbulent Mach numbers in excess of 0.2 are reached in some boundary layer experiments \citep[e.g.][]{Spina1987}, 
where the mean flows are typically too fast to enable high resolution measurements. 

Our experiments in the VDSSV (gray trapezoid) bridge a sparsely populated space between previous experiments that tend to occupy the low Mach number, high Reynolds number region, whereas numerical simulations reside in the high $M_t$, low $R_\lambda$ region. 
To the left and right, the space is bounded by the lowest (0.05\,bar)
and highest (15\,bar) gas pressures in the VDSSV. 
To the top right, the space is bounded by the power of the fan, where $M_{t} \sim R_{\lambda}^{-1}$ since the maximum power of the fan is $P_{max} \sim R_{\lambda}^{2}M_{t}^2$, which assumes $P_{max} \sim \rho U_{j}^{3}$ and $u \sim U_{j}$. 
In the VDSSV, $M_{t}$ and $R_{\lambda}$ can be adjusted independently by changing between working gases that have different sound speeds, and by adjusting mean jet velocities (moving up and down along the blue and red curves).
Two representative trajectories 
followed by our experiments for 1\,bar air and 0.165\,bar SF\textsubscript{6}
show that the Mach number grows approximately as the square of the Reynolds number (black line: $M_t \sim R_\lambda^2$) while increasing the fan speed and holding all else constant. 
This quadratic scaling holds if the flow velocity changes while its length scales do not, 
since $M_t \sim u^{\prime}$ and $R_\lambda \sim u^{\prime 1/2}$. 

\subsection{Jet profile characteristics}
The jet at its center is shear free 
and exhibits velocity fluctuations up to 15\,m/s. 
Velocity measurements at station 1 were performed to assess the flow structure of the jet in its initial development.
Turbulence measurements were performed in the developed region of the jet 
and past a wake region that extends to approximately $x/D_f$ = 5, 
where shear layers emanating from the fan become well-mixed with the wake 
and form a jet with an approximately self-similar profile. 

\subsubsection{Similarity at station 1}

\begin{figure}
\centering
	        \includegraphics[width=0.5\textwidth, trim=0cm 0cm 0cm 0cm,  clip]{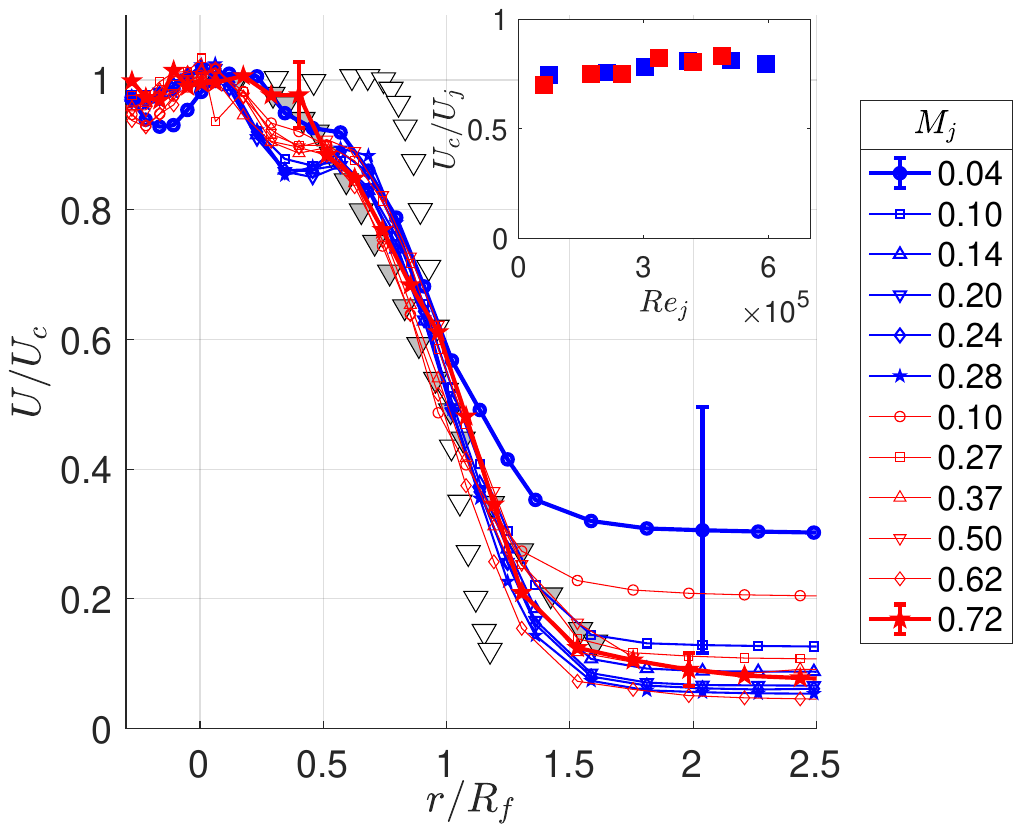}
	        \caption{
	        Radial mean velocity profiles close to the outlet of the fan ($x/D_f = 1.9$) 
	        normalized by the centerline mean velocity ($U_c$) 
	        in air (blue) and SF\textsubscript{6} (red)
	        are approximately invariant with respect to changes in Mach number. 
            The radius of the fan is $R_{f} = D_{f}/2$ = 45\,mm.
            At large $r$ the differences between the measured profiles and zero 
            are not significant since 
            the uncertainty is large at small velocities. 
	        Data from \citet{Narayanan2002} at $M_{j} = 0.6$ 
	        (triangles, open and closed symbols at $x/D_f$ = 1 and 4, respectively) 
	        are shown for reference,
	        for which $R_f$ is the half-width of the jet ($R_{1/2}$) 
	        and not the radius of the jet orifice. 
	        {\it Inset: } Centerline mean velocity, 
	        $U_c/U_j$, as a function of jet Reynolds number, $Re_j$ = $U_j D_f/\nu$. 
	        }
	        \label{fig:4}
\end{figure}

At $x/D_f$ = 1.9, 
near the fan outlet, 
the mean velocity profiles were approximately invariant with respect to changes in the Mach number  (Fig.~\ref{fig:4}). 
The profiles were measured at six representative pairs of jet Mach numbers 
ranging from $M_j$ = 0.04 to 0.72, 
with each pair at a constant Reynolds number. 
There is a velocity deficit in the central part of the jet out to 0.5$R_{f}$ from the centerline, 
which corresponds to a wake whose width is approximately the same as the fan motor diameter. 
Normalization of radial positions by the fan radius collapses all of the data out to at least $r/R_f \approx$ 1, consistent with the width being determined by the size of the orifice near the outlet. At larger radial distances, the seemingly higher velocities outside of the jet core at the two lowest $M_{j}$ are a result of a bias in the measurement due to two reasons. First, since the hot-wire measurements are unsigned, the resulting error is positive and thus cumulative. Second, in addition to random error we have systematic errors due to deviations of the calibration curve detailed in Section \ref{sec:experimental method} from King's Law at very low speeds. Error bars at the highest and lowest $M_{j}$ indicate that the differences between the speeds at the tails are statistically insignificant. 
The inset shows the jet centerline velocity, normalized by the jet exit velocity, as a function of jet Reynolds number, which approaches a constant value of approximately 0.8, a lower value compared to a compressible jet in \citet{Narayanan2002} due to the influence of the wake. 

Data from \citet{Narayanan2002} at $M_{j}$ = 0.6 are shown for comparison with our data, which displays a top-hat profile characteristic of near-orifice jet profiles at $x/D$ = 1 that quickly decays toward a rounded shape at $x/D$ = 4. 
The profiles in our experiment at two diameters are bounded by the data from \citet{Narayanan2002} for $r/R_{f} >$ 0.5, but presents flow irregularities at smaller radial distances due to the influence of the fan geometry.

\subsubsection{Similarity at station 2}

\begin{figure}
    \centering
    \includegraphics[width=0.5\textwidth, trim=0cm 0cm 0cm 0cm,  clip]{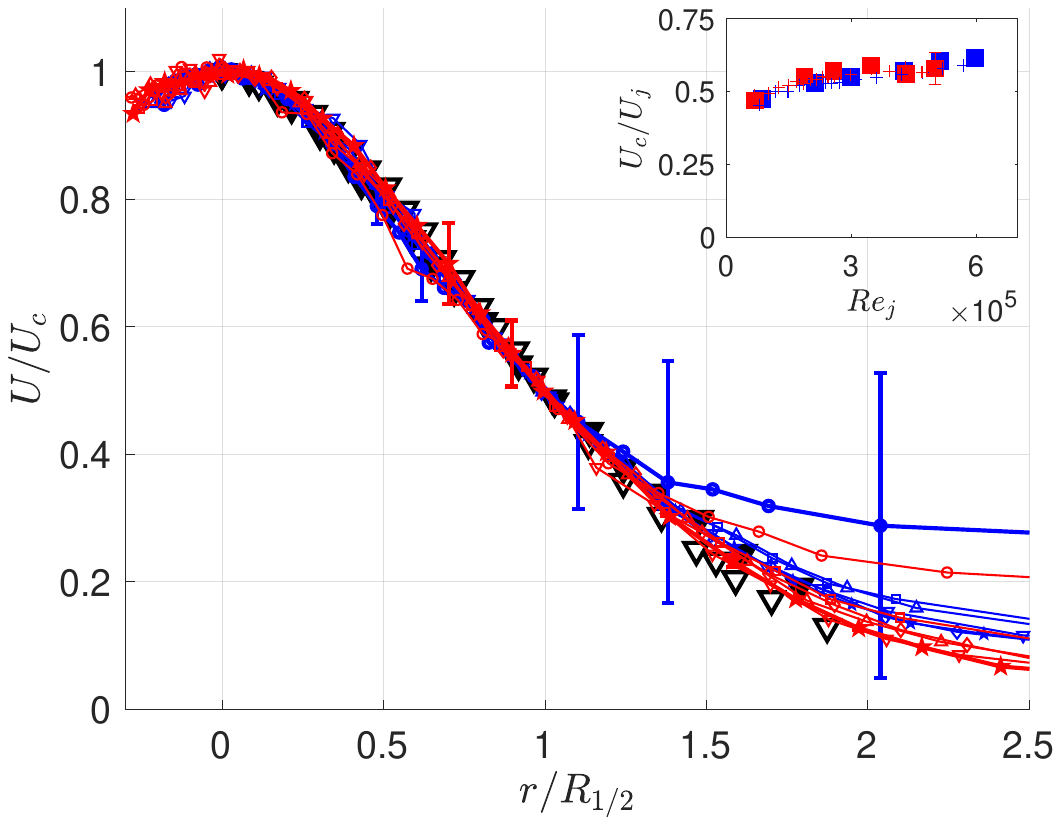}	
	\caption{
	The profiles collapse when scaled by their half-widths, $R_{1/2}$. 
    Data from \citet{Narayanan2002} at $M_{j}$ = 0.6 (black triangles) are shown for reference,
    with open and closed symbols 
    corresponding to streamwise positions $x/D_f$ = 6 and 10 respectively. 
	The error bars represent uncertainty in the calibration of the hot wire, 
	which is largest at low speeds and far from the jet axis. 
    Different markers correspond to Mach numbers 
    as in the legend for Fig.~\ref{fig:4}. 
    \textit{Inset:} 
    Jet centerline speed, $U_c/U_j$, variation with jet Mach number 
    with error bars representing uncertainty in calibration 
    (many of them smaller than the symbols). 
    }
    \label{fig:5}
\end{figure}

At 9.1 diameters from the fan (Fig.~\ref{fig:5}) 
the data collapse with each other 
when radial positions are normalized by the half-width radii of the jets ($R_{1/2}$). 
The profiles align with those from \citet{Narayanan2002} at 
similar Mach numbers and distances from the orifice ($M_{j} =$ 0.6 and $x/D =$ 6 and 10), 
and are indistinguishable from the normalized profiles of an incompressible jet at 98 diameters from the orifice (not shown) in \citet{Wygnanski1969}. 
These results suggest an approximate self-similarity of the jet. 

\subsubsection{Jet width development}

\begin{figure}
    \centering
	\includegraphics[width = 0.5\textwidth,trim=0cm 0cm 0cm 0cm, clip]{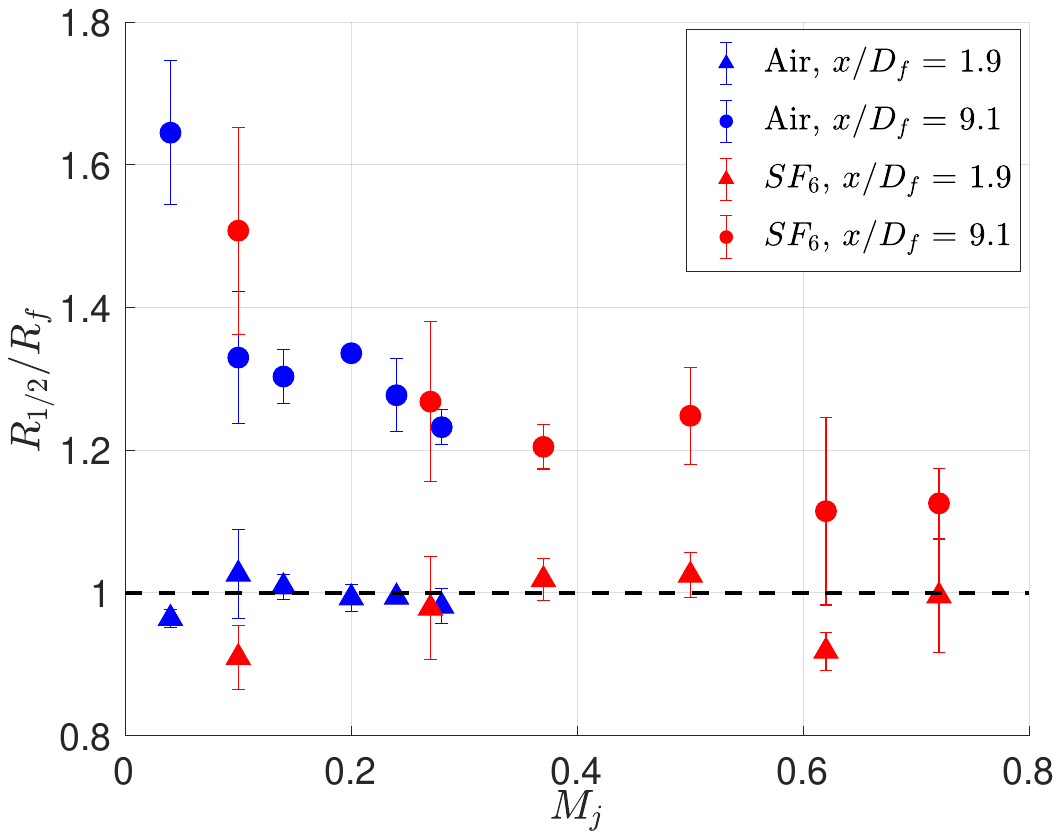}
	\includegraphics[width = 0.5\textwidth,trim=0cm 0cm 0cm 0cm, clip]{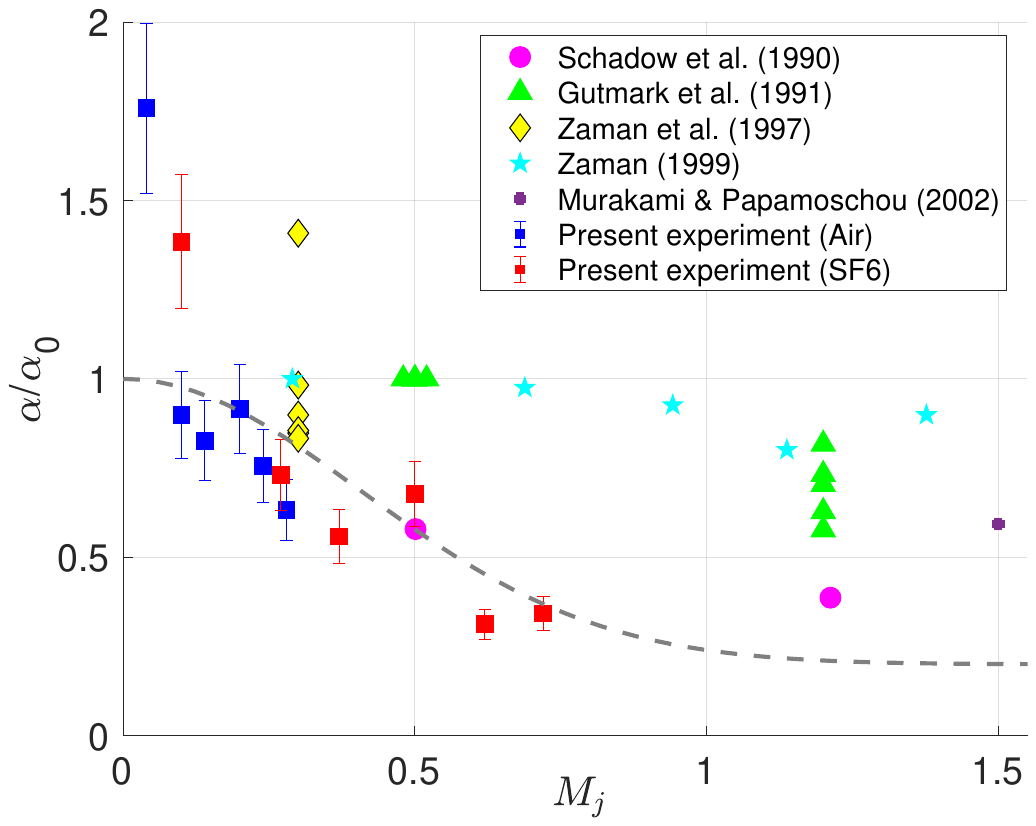}
	\caption{
	\textit{Top:} Jet widths at $x/D_f$ = 9.1 (circles)
	decrease with increasing Mach number, 
	while at station 1 (triangles) the width does not change. 
    The error bars are the differences between two experiments performed under the same conditions. 
	\textit{Bottom:} We interpret the difference between down- and up-stream jet widths 
	as a spreading rate, 
	$\alpha$. 
	This spreading rate 
	decreases with increasing jet Mach number, $M_j$. 
    The data are qualitatively consistent with a Langley-like curve, embodied in a fit (dashed black line) of the form 
    $a + (1-a) \exp{(-b M_{j}^{2})}$ 
    where $a$ and $b$  
    are equal to 1/5  
    and 3/2, respectively. 
	}
    \label{fig:6}
\end{figure}

At low Mach numbers, the jet spreads such that its diameter 
is at least one-third larger at 9 diameters downstream than at 2 diameters downstream 
(Fig.~\ref{fig:6}, top). 
As the Mach number increases, 
the width of the jet at station 2 (circles) narrows 
and approaches the jet width at station 1 (triangles), 
and so the width of the orifice itself. 
In other words, the jet approaches a ``turbulent laser''  
that does not diverge with increasing distance from its source at high Mach numbers.
Axial momentum transport may be inhibited in the radial direction 
at higher Mach numbers by mechanisms including 
smaller transverse velocity fluctuations, which are not measured here \citep{Arun2019,matsunolele2021}. 

We characterize the jet spreading rate by $\alpha \equiv$ $\Delta R / \Delta x$, which is the difference between up- and down-stream diameters, $\Delta R$ = $ (R_{1/2}(x/D_f = 9.1) - R_{1/2}(x/D_f = 1.9))$, divided by the distance between the up- and down-stream positions, $\Delta x$. 
We make the spreading rate relative to the median of the ones for which $M_j$ < 0.2, which is $\alpha_0$,
and we assumed a constant virtual origin with increasing Reynolds number. 
The virtual origin is insensitive to the Reynolds number, 
but depends strongly on inflow conditions and strength of disturbances \citep{pitts1991reynolds,boersma1998numerical}. 
Normalized jet growth rates decrease with jet Mach number (Fig.~\ref{fig:6}, bottom). This trend is qualitatively consistent with and analogous to the Langley curve (black dashed line), where a $40\%$ decrease in normalized compressible shear layer growth rate can be observed at a jet Mach number as low as 0.5. 
Our data is compared to previous 
jet studies that measured spreading rates in the range of axial positions of $2<x/D<30$ \citep{Schadow1990,Gutmark1991} or $2<x/D<20$ \citep{Zaman1997,Zaman1999,Murakami2002}, which overlaps with the distances over which spreading rates were characterized in the present experiment.
The apparent initial increase in spreading rate is due to a high uncertainty on the fringes of the jet at the lowest $M_{j}$. Unlike the relatively good collapse in the mixing layer study by \citet{barone2006validation}, 
the variability in the jet data for any given Mach number is an indication of sensitivity to initial and boundary conditions, and therefore of non-universality in terms of the Mach number, in addition to measurement uncertainties and differences in practical definitions of the spreading rate between experiments. 

\subsection{Turbulence at station 2}

Here we examine the outer scales of the turbulence at station 2, 
located at the furthest distance from the orifice 
at which we measured in the present experiments ($x/D_f = 9$). 

\begin{figure}
    \centering
    \includegraphics[width=0.5 \textwidth,trim=0cm 0cm 0cm 0cm,  clip]{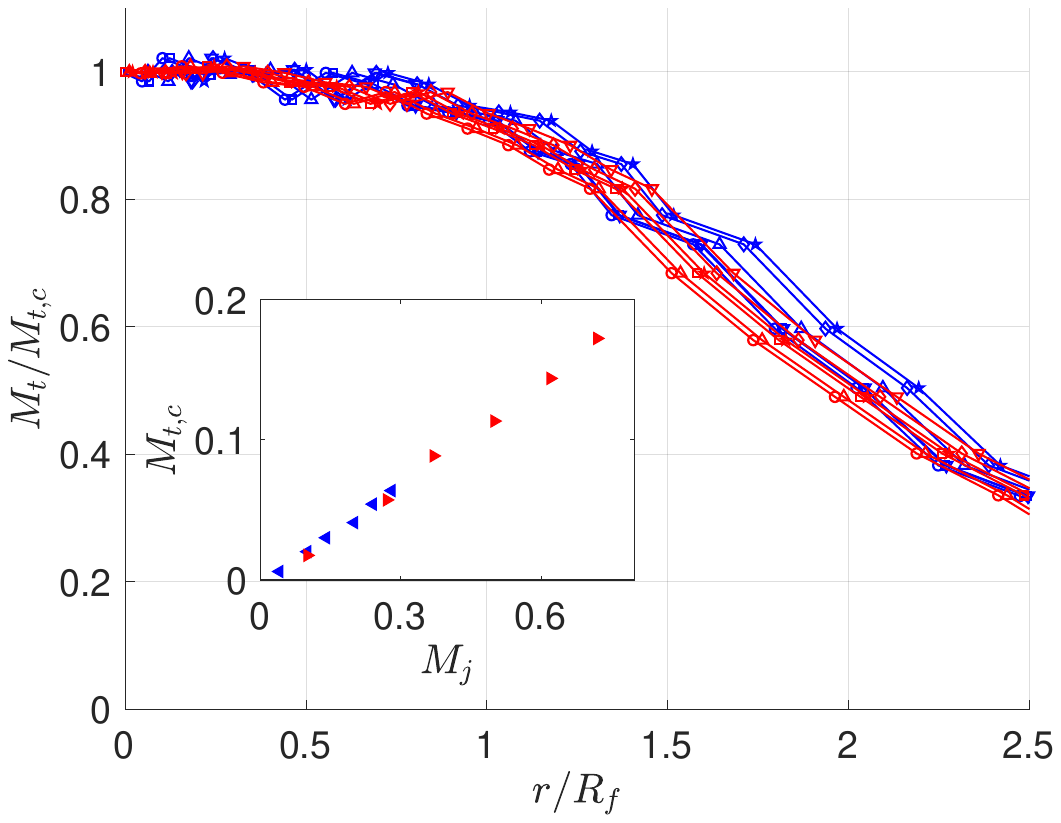}
\caption{
    From the centerline out to about one fan radius ($r/R_f$ = 1), 
    the turbulent Mach number ($M_{t}$) 
    is approximately homogeneous 
    in air (blue) and SF\textsubscript{6} (red) at $x/D_f = 9.1$. 
    \textit{Inset:} 
    The turbulence Mach number on the centerline, $M_{t,c}$, 
    increases linearly with the jet Mach number, $M_j$. 
    Markers correspond to Mach numbers as in the legend for  Fig.~\ref{fig:4}. 
    }
    \label{fig:7}
\end{figure}

\subsubsection{Homogeneity}
At station 2 ($x/D_f$ = 9.1), 
the turbulence in the jet is approximately homogeneous within one jet diameter, 
in the sense that the turbulent Mach number 
is constant to within about 5$\%$ up to one fan radius for all conditions
(Fig.~\ref{fig:7}). 
It is interesting that the turbulent Mach number profiles do not narrow as quickly with Mach number as the mean profiles, so that the turbulence intensity, $u^{\prime}/U$, 
grows to larger values at the fringes of the jet at higher Mach numbers. 
The inset shows that the turbulent Mach number at the center of the jet increases linearly with the jet Mach number, 
attaining a maximum value of $M_{t}$ = 0.17 when $M_{j} \approx$ 0.7. 
On the center-line, 
the turbulence intensity at station 2 was between 20$\%$ and 24$\%$ of the mean velocity. 
This compares with upstream intensities between 8$\%$ and 14$\%$ at $x/D_f$ = 1.9 
and on the center-line. 

\begin{figure}
	\centering
	\includegraphics[width=0.5 \textwidth,trim=0cm 0cm 0cm 0cm, clip]{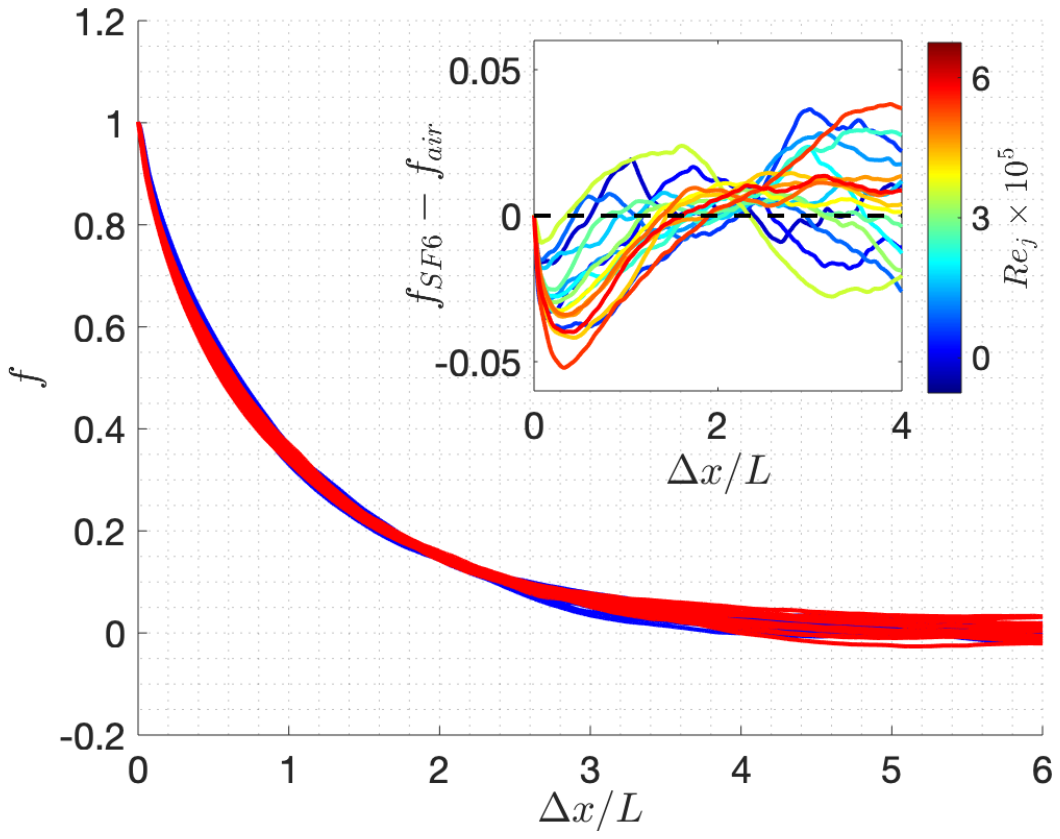}
	\includegraphics[width=0.5 \textwidth,trim=0cm 0cm 0cm 0cm,clip]{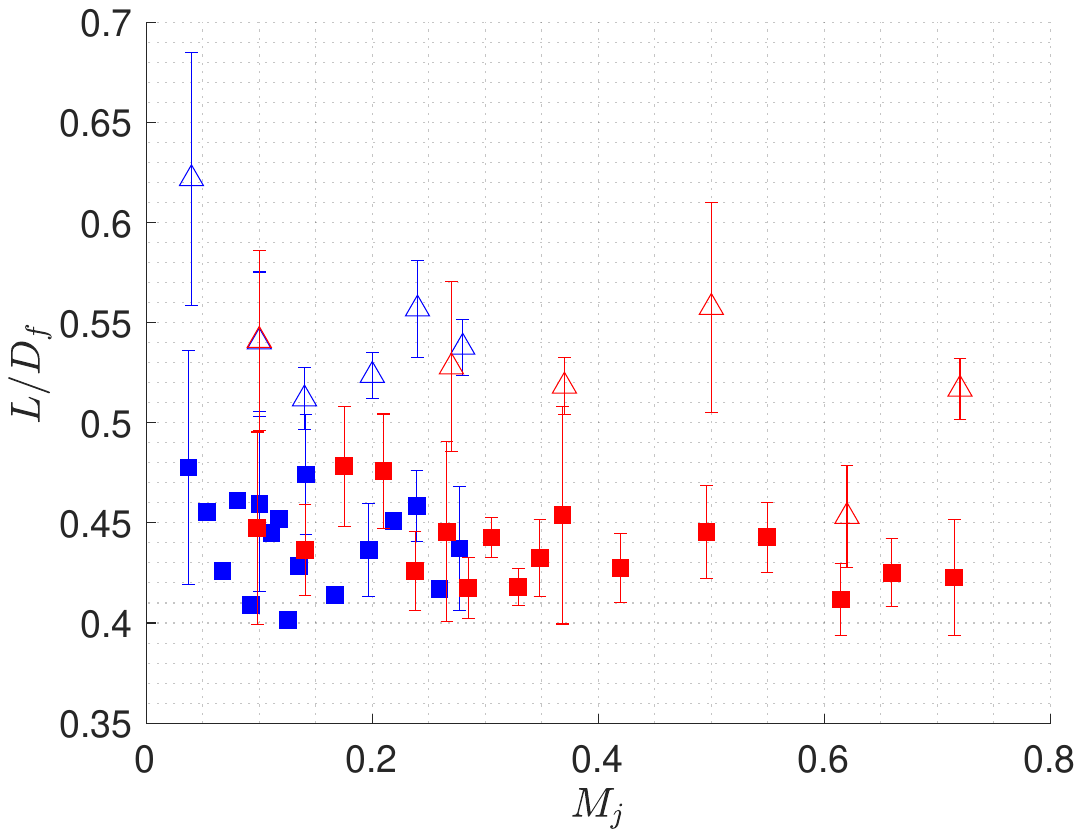}
	\caption{
	\textit{Top:} The velocity correlation functions along the centerline at $x/D_{f}$ = 9.1 are approximately self-similar 
	with respect to changes in Reynolds and Mach numbers, 
	with a small systematic variation shown in the inset. 
	The data are plotted as a function of separation, $\Delta x$, 
	normalized by their integral time scales, $L$. 
	{\it Inset:} 
	The differences between pairs of correlation functions measured in SF$_6$ and air 
	at approximately the same Reynolds number, where 
	the Reynolds number increases from blue to red curves. 
	\textit{Bottom:} 
	The integral length scale, $L$, 
    in fan diameters, $D_{f}$,
    is approximately independent of the Mach number on the center-line (closed symbols). 
    Length scales at the margin of the jet (open symbols) 
    are about 20\% larger than those on the center-line. 
    The error bars represent variation between 
    different definitions of $L$, including $L_{e}$, 
    which is the distance at which the autocorrelation function is equal to $1/e$. 
	}
	\label{fig:8}
\end{figure}

\subsubsection{Autocorrelations and integral scales}
Along the centerline and at station 2, 
the longitudinal velocity correlation functions, $f$,  
collapse approximately when rescaled by the correlation lengths, $L$, 
defined in Sec.~\ref{sec:experimental method} (Fig.~\ref{fig:8}, top). 
These correlation lengths, $L$, are themselves approximately 
independent of the Mach number (Fig.~\ref{fig:8}, bottom). 
Despite the strong fluctuations, 
we interpret the correlations as spatial functions of $\Delta x = U \Delta t$, 
according to Taylor's Hypothesis \citep{Lee1992}, 
where $\Delta t$ is the time interval between velocity measurements, 
though this interpretation is not necessary here. 

A small but systematic trend in the correlation functions 
is magnified by making the difference 
between the high (SF$_6$) and low (air) Mach number correlation functions 
at each Reynolds number (Fig.~\ref{fig:8}, inset). 
When the Mach number is raised at constant Reynolds number, 
     slightly faster decorrelations are introduced at separations smaller than the integral length scale, as quantified by the magnitude of the first troughs in the inset, and are compensated by stronger correlations at large separations. This effect is more prominent with increasing Reynolds number (indicated by the colorbar). 

In incompressible turbulence, integral scales depend weakly on the Reynolds number. 
The dependence on Mach number, however, is not entirely clear. 
At $x/D_f$ = 9.1 and the jet center,
integral scales normalized by the fan diameter $D_{f}$ 
remain relatively constant with jet Mach number at about $0.45D_{f}$, 
closely following the fan radius (Fig.~\ref{fig:8}, bottom). 
This aligns with the notion that integral scales are geometrically related to the scales of turbulence energy injection \citep{White2002}. 
Integral scales are slightly larger at the jet margins ($r=R_{f}$) than at the jet center, 
achieving values slightly in excess of the fan radius on average. 
\subsubsection{Energy spectra}

\begin{figure}
\centering
	        \includegraphics[width=0.5\textwidth,trim=0cm 0cm 0cm 0cm, clip]{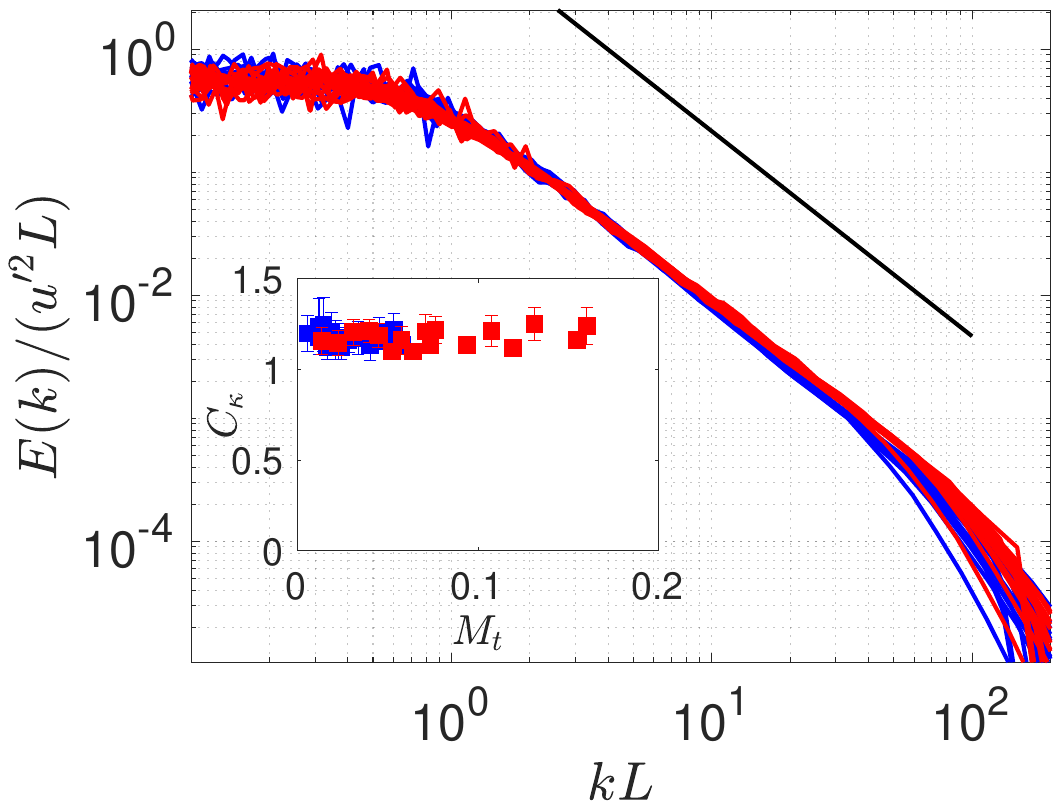} 
	        \caption{
	        Energy spectra in air (blue) and SF\textsubscript{6} (red) 
	        collapse at small wavenumbers, $k$, under normalization by $u^\prime$ and $L$. 
	        The solid black line is a -5/3$^\mathrm{rds}$ power law. 
	        \textit{Inset:} 
	        The Kolmogorov constant, $C_{\kappa}$, for these spectra 
	        is approximately independent of the Mach number, $M_t$. }
	        \label{fig:9}
\end{figure}

We observe that all spectra display a -5/3$^\mathrm{rds}$ scaling region  consistent with incompressible dynamics
up to the highest turbulent Mach number we reached, $M_{t}$ = 0.17 (Fig.~\ref{fig:9}). 
The observed inertial subrange extends over one to two decades for all flow conditions, 
indicating well-developed turbulence. 
At increasing flow speeds, 
our probes spatially truncated the smallest scales and did not capture them at the highest $R_{\lambda}$ and $M_{t}$. 
At large wavenumbers, the cutoff in the spectra are determined by filters and not by the dissipation for all but the lowest Reynolds number data. The corresponding Kolmogorov scales range from 20\,$\mu m$ to 100\,$\mu m$ or between 400 and 2000 integral scales ($L$).
The inset shows that the Kolmogorov constant, $C_{\kappa}$, is independent of $M_{t}$. $C_{\kappa}$ was estimated assuming isotropy, a -5/3$^{rds}$ scaling between $kL$ = 2 and 20, and a constant value of one for $C_{\epsilon} = \varepsilon L/u^{\prime3}$. Since we did not independently measure the dissipation, it is the trend, and not the particular value of $C_{\kappa}$, to which we draw attention. 
\section{Summary and conclusion}

We detail the main design principles and capabilities of the VDSSV, 
a new facility that generates high-speed, subsonic compressible turbulent flows. 
Using hot-wire anemometry capable of resolving inertial subrange statistics, 
we report the first quantitative measurements in locally isotropic compressible turbulence. 
We highlight the VDSSV's unique ability to raise the Mach number at constant Reynolds number by switching the gas composition, 
while also attaining the highest turbulent Mach number, $M_{t} = 0.17$, 
in a free shear flow and in a controlled laboratory setting, to the best of our knowledge. 
We quantify turbulence at high Mach numbers at flow speeds low enough to enable high resolution measurements. 
Compressible flows in the pressure vessel were produced by means of a jet whose outlet is turbulent
in SF\textsubscript{6}. 
The jet's spatial development showed that downstream mean profiles were approximately self-similar, while also spreading at slower rates at high Mach numbers. 
The pressure vessel can easily be extended to access larger $x/D_f$ with the addition of cylindrical sections to its middle. Other upgrades may include the addition of heaters to generate a hot jet discharging into a colder environment.
The streamwise turbulence in the jet was homogeneous in the radial direction within 5$\%$ up to one jet radius. 
Autocorrelation differences at a constant Reynolds but different Mach number revealed that initially faster decorrelations in the high Mach number cases are compensated by stronger correlations at large distances. 
Further experimentation is required to establish a clear link between these decorrelations and compressible or acoustic motions. 
Spectra and their Kolmogorov constants were unchanged by increases in the Mach number. 
Future experiments will aim to further characterize the jet at a broader range of streamwise positions, 
to provide more accurate estimates for the growth rates, and to quantify the recirculation on the sides of the jet. 
These experiments are projected to reach turbulent Mach numbers as high as 0.4 with the use of active grids \citep{Griffin2019} and will additionally use optical instruments to track particles \citep{kearney2020} and to measure density fluctuations. 

\begin{acknowledgements}
The authors are grateful 
to N.\,Dam and W.\,van de Water for the pressure vessel 
and for assistance from 
D.\,Donzis, L.\,Mydlarski and M.\,Ulinski. 
The authors also thank 
J.\,John, 
E.\,Liu, 
and H.\,Rivera 
for helpful discussions, 
a team of Master's students including 
E.\,Bair, 
A.\,Berberian, 
D.\,Cohen, 
D.\,Feng, 
T.\,Green, 
B.\,Oster, and 
K.\,Rajsky for designing, manufacturing and testing elements of the facility, 
and a team of undergraduates including 
Y.\,Atiq, 
S.\,Bell, 
W.\,Chan, 
M.\,Chen, 
S.\,DePue, 
C.\,Kartawira, 
A.\,Ramos Figueroa, 
K.\,Roberts, and 
C.\,Vahn for their help with data collection and experimental setup. 
\end{acknowledgements}



\bibliographystyle{spbasic}      
\bibliography{biblio_1}   


\end{document}